\documentclass[longauth]{aa} 
\usepackage{graphicx}
\usepackage{txfonts}
\usepackage{xspace}
\usepackage{natbib}
\usepackage{url}
\usepackage{color}
\usepackage{hyperref}

\newcommand{\percm}{\,cm$^{-2}$}	
\newcommand{\pers}{\,s$^{-1}$}	
\newcommand{\fluxcgs}{erg~cm$^{-2}$~s$^{-1}$}
\newcommand{\ixpe}{\textit{IXPE}}
\newcommand{\nustar}{\textit{NuSTAR}}

\newcommand{\xspec}{\textsc{xspec}}
\newcommand{\diskbb}{\texttt{diskbb}}
\newcommand{\comptt}{\texttt{comptt}}
\newcommand{\comptb}{\texttt{comptb}}
\newcommand{\relxillns}{\texttt{relxillns}}
\newcommand{\polconst}{\texttt{polconst}}
\newcommand{\tbabs}{\texttt{tbabs}}
\newcommand{\relline}{\texttt{relline}}

\begin{document}

\title{X-ray polarimetry and spectroscopy of the neutron star low-mass X-ray binary GX 9+9: an in-depth study with \ixpe\ and \nustar}

\titlerunning{X-ray polarimetry and spectroscopy of GX 9+9}
\authorrunning{Ursini F. et al.} 

\author{
F.~Ursini\inst{\ref{in:UniRoma3}} \and 
R.~Farinelli\inst{\ref{in:INAF-OAS}} \and
A.~Gnarini\inst{\ref{in:UniRoma3}} \and
J.~Poutanen\inst{\ref{in:Turku}} \and 
S.~Bianchi\inst{\ref{in:UniRoma3}} \and
F.~Capitanio\inst{\ref{in:INAF-IAPS}} \and
A.~Di~Marco\inst{\ref{in:INAF-IAPS}} \and
S.~Fabiani\inst{\ref{in:INAF-IAPS}} \and
F.~La~Monaca\inst{\ref{in:INAF-IAPS}} \and
C.~Malacaria\inst{\ref{in:ISSI}} \and
G.~Matt\inst{\ref{in:UniRoma3}} \and
R.~Miku{\v s}incov\'a\inst{\ref{in:UniRoma3}} \and
M.~Cocchi\inst{\ref{in:INAF-OAC}} \and
P.~Kaaret\inst{\ref{in:NASA-MSFC}} \and
J.~J.~E.~Kajava\inst{\ref{in:Turku},\ref{in:Serco-ESA}} \and
M.~Pilia\inst{\ref{in:INAF-OAC}} \and
W.~Zhang\inst{\ref{in:NAO-CAS}}  
\and I. Agudo \inst{\ref{in:CSIC-IAA}}
\and L. A. Antonelli \inst{\ref{in:INAF-OAR},\ref{in:ASI-SSDC}} 
\and M. Bachetti \inst{\ref{in:INAF-OAC}} 
\and L. Baldini  \inst{\ref{in:INFN-PI},	\ref{in:UniPI}} 
\and W. H. Baumgartner  \inst{\ref{in:NASA-MSFC}} 
\and R. Bellazzini  \inst{\ref{in:INFN-PI}}  
\and S. D. Bongiorno \inst{\ref{in:NASA-MSFC}} 
\and R. Bonino  \inst{\ref{in:INFN-TO},\ref{in:UniTO}}
\and A. Brez  \inst{\ref{in:INFN-PI}} 
\and N. Bucciantini 
\inst{\ref{in:INAF-Arcetri},\ref{in:UniFI},\ref{in:INFN-FI}} 
\and S. Castellano \inst{\ref{in:INFN-PI}}  
\and E.	Cavazzuti \inst{\ref{in:ASI}} 
\and C.-T.	Chen \inst{\ref{in:USRA-MSFC}}
\and S. Ciprini \inst{\ref{in:INFN-Roma2},\ref{in:ASI-SSDC}}
\and E.	Costa \inst{\ref{in:INAF-IAPS}} 
\and A.	De Rosa \inst{\ref{in:INAF-IAPS}} 
\and E.	Del Monte \inst{\ref{in:INAF-IAPS}} 
\and L.	Di Gesu \inst{\ref{in:ASI}} 
\and N. Di Lalla \inst{\ref{in:Stanford}}
\and I.	Donnarumma \inst{\ref{in:ASI}}
\and V. Doroshenko \inst{\ref{in:Tub}}
\and M. Dov\v{c}iak \inst{\ref{in:CAS-ASU}}
\and S.	R. Ehlert \inst{\ref{in:NASA-MSFC}}  
\and T. Enoto \inst{\ref{in:RIKEN}}
\and Y. Evangelista \inst{\ref{in:INAF-IAPS}}
\and R. Ferrazzoli \inst{\ref{in:INAF-IAPS}} 
\and J.	A. Garcia \inst{\ref{in:Caltech}}
\and S. Gunji\inst{\ref{in:Yamagata}} 
\and K. Hayashida \inst{\ref{in:Osaka}}\thanks{Deceased.} 
\and J. Heyl \inst{\ref{in:UBC}}
\and W.	Iwakiri \inst{\ref{in:Chiba}} 
\and S. G. Jorstad \inst{\ref{in:BU},\ref{in:SPBU}} 
\and V. Karas \inst{\ref{in:CAS-ASU}}
\and F.	Kislat \inst{\ref{in:UNH}} 
\and T.	Kitaguchi  \inst{\ref{in:RIKEN}} 
\and J. J. Kolodziejczak \inst{\ref{in:NASA-MSFC}} 
\and H. Krawczynski  \inst{\ref{in:WUStL}}
\and L. Latronico  \inst{\ref{in:INFN-TO}} 
\and I. Liodakis \inst{\ref{in:FINCA}}
\and S.	Maldera \inst{\ref{in:INFN-TO}}  
\and A. Manfreda \inst{\ref{INFN-NA}}
\and F. Marin \inst{\ref{in:Strasbourg}} 
\and A.	Marinucci \inst{\ref{in:ASI}} 
\and A. P. Marscher \inst{\ref{in:BU}} 
\and H. L. Marshall \inst{\ref{in:MIT}}
\and F.	Massaro \inst{\ref{in:INFN-TO},\ref{in:UniTO}} 
\and I. Mitsuishi \inst{\ref{in:Nagoya}} 
\and T.	Mizuno \inst{\ref{in:Hiroshima}} 
\and F.	Muleri \inst{\ref{in:INAF-IAPS}} 
\and M. Negro \inst{\ref{in:UMBC},\ref{in:NASA-GSFC},\ref{in:CRESST}} 
\and C.-Y. Ng \inst{\ref{in:HKU}}
\and S. L.	O'Dell \inst{\ref{in:NASA-MSFC}}  
\and N.	Omodei \inst{\ref{in:Stanford}}
\and C.	Oppedisano \inst{\ref{in:INFN-TO}}  
\and A.	Papitto \inst{\ref{in:INAF-OAR}}
\and G.	G. Pavlov \inst{\ref{in:PSU}}
\and A. L. Peirson \inst{\ref{in:Stanford}}
\and M.	Perri \inst{\ref{in:ASI-SSDC},\ref{in:INAF-OAR}}
\and M. Pesce-Rollins \inst{\ref{in:INFN-PI}} 
\and P.-O.	Petrucci \inst{\ref{in:Grenoble}} 
\and M. Pilia \inst{\ref{in:INAF-OAC}} 
\and A.	Possenti \inst{\ref{in:INAF-OAC}} 
\and S.	Puccetti \inst{\ref{in:ASI-SSDC}}
\and B. D. Ramsey \inst{\ref{in:NASA-MSFC}}  
\and J. Rankin \inst{\ref{in:INAF-IAPS}} 
\and A. Ratheesh \inst{\ref{in:INAF-IAPS}} 
\and O.	J. Roberts \inst{\ref{in:USRA-MSFC}}
\and R. W. Romani \inst{\ref{in:Stanford}}
\and C. Sgr\`o \inst{\ref{in:INFN-PI}}  
\and P. Slane \inst{\ref{in:CfA}}  
\and P. Soffitta \inst{\ref{in:INAF-IAPS}} 
\and G.	Spandre \inst{\ref{in:INFN-PI}} 
\and D. A.	Swartz \inst{\ref{in:USRA-MSFC}}
\and T. Tamagawa \inst{\ref{in:RIKEN}}
\and F. Tavecchio \inst{\ref{in:INAF-OAB}}
\and R. Taverna \inst{\ref{in:UniPD}} 
\and Y.	Tawara \inst{\ref{in:Nagoya}}
\and A. F. Tennant \inst{\ref{in:NASA-MSFC}}  
\and N. E. Thomas \inst{\ref{in:NASA-MSFC}}  
\and F.	Tombesi  \inst{\ref{in:UniRoma2},\ref{in:INFN-Roma2},\ref{in:UMd}}
\and A. Trois \inst{\ref{in:INAF-OAC}}
\and S. S. Tsygankov \inst{\ref{in:Turku}}
\and R. Turolla \inst{\ref{in:UniPD},\ref{in:MSSL}}
\and J. Vink \inst{\ref{in:Amsterdam}}
\and M. C. Weisskopf \inst{\ref{in:NASA-MSFC}} 
\and K.	Wu \inst{\ref{in:MSSL}}
\and F. Xie \inst{\ref{in:GSU},\ref{in:INAF-IAPS}}
\and S. Zane  \inst{\ref{in:MSSL}}
}

\institute{
Dipartimento di Matematica e Fisica, Universit\`a degli Studi Roma Tre, via della Vasca Navale 84, 00146 Roma, Italy \label{in:UniRoma3} 
\and
INAF -- Osservatorio di Astrofisica e Scienza dello Spazio di Bologna, Via P. Gobetti 101, I-40129 Bologna, Italy \label{in:INAF-OAS} 
\and
Department of Physics and Astronomy, FI-20014 University of Turku, Finland \label{in:Turku} 
\and
{\fontdimen2\font=0.9\fontdimen2\font INAF -- IAPS, via del Fosso del Cavaliere 100, I-00113 Roma, Italy} \label{in:INAF-IAPS} 
\and
International Space Science Institute, Hallerstrasse 6, 3012 Bern, Switzerland \label{in:ISSI} 
\and
INAF -- Osservatorio Astronomico di Cagliari, via della Scienza 5, I-09047 Selargius (CA), Italy \label{in:INAF-OAC} 
\and
NASA Marshall Space Flight Center, Huntsville, AL 35812, USA \label{in:NASA-MSFC} 
\and
Serco for the European Space Agency (ESA), European Space Astronomy Centre, Camino Bajo del Castillo s/n, E-28692 Villanueva de la Ca\~{n}ada, Madrid, Spain \label{in:Serco-ESA} 
\and
National Astronomical Observatories, Chinese Academy of Sciences, 20A Datun Road, Beijing 100101, China \label{in:NAO-CAS}
\and 
Instituto de Astrof\'{i}sicade Andaluc\'{i}a -- CSIC, Glorieta de la Astronom\'{i}a s/n, 18008 Granada, Spain \label{in:CSIC-IAA}
\and 
INAF Osservatorio Astronomico di Roma, Via Frascati 33, 00040 Monte Porzio Catone (RM), Italy \label{in:INAF-OAR} 
\and 
Space Science Data Center, Agenzia Spaziale Italiana, Via del Politecnico snc, 00133 Roma, Italy \label{in:ASI-SSDC}
\and 
Istituto Nazionale di Fisica Nucleare, Sezione di Pisa, Largo B. Pontecorvo 3, 56127 Pisa, Italy \label{in:INFN-PI}
\and  
Dipartimento di Fisica, Universit\`{a} di Pisa, Largo B. Pontecorvo 3, 56127 Pisa, Italy \label{in:UniPI} 
\and  
Istituto Nazionale di Fisica Nucleare, Sezione di Torino, Via Pietro Giuria 1, 10125 Torino, Italy  \label{in:INFN-TO}	
\and  
Dipartimento di Fisica, Universit\`{a} degli Studi di Torino, Via Pietro Giuria 1, 10125 Torino, Italy \label{in:UniTO} 
\and   
INAF Osservatorio Astrofisico di Arcetri, Largo Enrico Fermi 5, 50125 Firenze, Italy 
\label{in:INAF-Arcetri} 
\and  
Dipartimento di Fisica e Astronomia, Universit\`{a} degli Studi di Firenze, Via Sansone 1, 50019 Sesto Fiorentino (FI), Italy \label{in:UniFI} 
\and   
Istituto Nazionale di Fisica Nucleare, Sezione di Firenze, Via Sansone 1, 50019 Sesto Fiorentino (FI), Italy \label{in:INFN-FI}
\and 
Agenzia Spaziale Italiana, Via del Politecnico snc, 00133 Roma, Italy \label{in:ASI}
\and 
Science and Technology Institute, Universities Space Research Association, Huntsville, AL 35805, USA \label{in:USRA-MSFC}
\and 
Istituto Nazionale di Fisica Nucleare, Sezione di Roma ``Tor Vergata'', Via della Ricerca Scientifica 1, 00133 Roma, Italy 
\label{in:INFN-Roma2}
\and 
Department of Physics and Kavli Institute for Particle Astrophysics and Cosmology, Stanford University, Stanford, California 94305, USA  \label{in:Stanford}
\and
Institut f\"ur Astronomie und Astrophysik, Universit\"at T\"ubingen, Sand 1, D-72076 T\"ubingen, Germany \label{in:Tub}
\and 
Astronomical Institute of the Czech Academy of Sciences, Boční II 1401/1, 14100 Praha 4, Czech Republic \label{in:CAS-ASU}
\and 
RIKEN Cluster for Pioneering Research, 2-1 Hirosawa, Wako, Saitama 351-0198, Japan \label{in:RIKEN}
\and 
California Institute of Technology, Pasadena, CA 91125, USA \label{in:Caltech}
\and 
Yamagata University,1-4-12 Kojirakawa-machi, Yamagata-shi 990-8560, Japan \label{in:Yamagata}
\and 
Osaka University, 1-1 Yamadaoka, Suita, Osaka 565-0871, Japan \label{in:Osaka}
\and 
University of British Columbia, Vancouver, BC V6T 1Z4, Canada \label{in:UBC}
\and 
International Center for Hadron Astrophysics, Chiba University, Chiba 263-8522, Japan \label{in:Chiba}
\and
Institute for Astrophysical Research, Boston University, 725 Commonwealth Avenue, Boston, MA 02215, USA \label{in:BU} 
\and 
Department of Astrophysics, St. Petersburg State University, Universitetsky pr. 28, Petrodvoretz, 198504 St. Petersburg, Russia \label{in:SPBU} 
\and 
Department of Physics and Astronomy and Space Science Center, University of New Hampshire, Durham, NH 03824, USA \label{in:UNH} 
\and 
Physics Department and McDonnell Center for the Space Sciences, Washington University in St. Louis, St. Louis, MO 63130, USA \label{in:WUStL}
\and 
Finnish Centre for Astronomy with ESO,  20014 University of Turku, Finland \label{in:FINCA}
\and 
Istituto Nazionale di Fisica Nucleare, Sezione di Napoli, Strada Comunale Cinthia, 80126 Napoli, Italy \label{INFN-NA}
\and 
Universit\'{e} de Strasbourg, CNRS, Observatoire Astronomique de Strasbourg, UMR 7550, 67000 Strasbourg, France \label{in:Strasbourg}
\and 
MIT Kavli Institute for Astrophysics and Space Research, Massachusetts Institute of Technology, 77 Massachusetts Avenue, Cambridge, MA 02139, USA \label{in:MIT}
\and 
Graduate School of Science, Division of Particle and Astrophysical Science, Nagoya University, Furo-cho, Chikusa-ku, Nagoya, Aichi 464-8602, Japan \label{in:Nagoya}
\and 
{\fontdimen2\font=0.8\fontdimen2\font Hiroshima Astrophysical Science Center, Hiroshima University, 1-3-1 Kagamiyama, Higashi-Hiroshima, Hiroshima 739-8526, Japan} \label{in:Hiroshima}
\and
University of Maryland, Baltimore County, Baltimore, MD 21250, USA \label{in:UMBC}
\and 
NASA Goddard Space Flight Center, Greenbelt, MD 20771, USA  \label{in:NASA-GSFC}
\and 
Center for Research and Exploration in Space Science and Technology, NASA/GSFC, Greenbelt, MD 20771, USA  \label{in:CRESST}
\and 
Department of Physics, University of Hong Kong, Pokfulam, Hong Kong \label{in:HKU}
\and 
Department of Astronomy and Astrophysics, Pennsylvania State University, University Park, PA 16801, USA \label{in:PSU}
\and 
Universit\'{e} Grenoble Alpes, CNRS, IPAG, 38000 Grenoble, France \label{in:Grenoble}
\and 
Center for Astrophysics, Harvard \& Smithsonian, 60 Garden St, Cambridge, MA 02138, USA \label{in:CfA} 
\and 
INAF Osservatorio Astronomico di Brera, via E. Bianchi 46, 23807 Merate (LC), Italy \label{in:INAF-OAB}
\and 
Dipartimento di Fisica e Astronomia, Universit\`{a} degli Studi di Padova, Via Marzolo 8, 35131 Padova, Italy \label{in:UniPD}
\and
Dipartimento di Fisica, Universit\`{a} degli Studi di Roma ``Tor Vergata'', Via della Ricerca Scientifica 1, 00133 Roma, Italy \label{in:UniRoma2}
\and
Department of Astronomy, University of Maryland, College Park, Maryland 20742, USA \label{in:UMd}
\and 
Mullard Space Science Laboratory, University College London, Holmbury St Mary, Dorking, Surrey RH5 6NT, UK \label{in:MSSL}
\and 
Anton Pannekoek Institute for Astronomy \& GRAPPA, University of Amsterdam, Science Park 904, 1098 XH Amsterdam, The Netherlands  \label{in:Amsterdam}
\and 
Guangxi Key Laboratory for Relativistic Astrophysics, School of Physical Science and Technology, Guangxi University, Nanning 530004, China \label{in:GSU}
}

\date{Accepted... Received...}

\abstract{
We report on a comprehensive analysis of simultaneous X-ray polarimetric and spectral data of the bright atoll source GX~9+9 with the Imaging X-ray Polarimetry Explorer (\ixpe) and \nustar. The source is significantly polarized in the 4--8 keV band, with a degree of $2.2\% \pm 0.5\%$ (uncertainty at the 68\% confidence level). The \nustar\ broad-band spectrum clearly shows an iron line, and is well described by a model including thermal disk emission, a Comptonized component, and reflection. From a spectro-polarimetric fit, we obtain an upper limit to the polarization degree of the disk of 4\% (at 99\% confidence level), while the contribution of Comptonized and reflected radiation cannot be conclusively separated. However, the polarization is consistent with resulting from a combination of Comptonization in a boundary or spreading layer, plus reflection off the disc, which gives a significant contribution in any realistic scenario. 
}


\maketitle



\section{Introduction}

Weakly magnetized neutron star low-mass X-ray binaries (NS-LMXBs) accrete mass via Roche lobe overflow from a low-mass stellar companion. They are historically divided into two classes, Z and atoll, according to their tracks on the colour-colour-diagram \citep{hasinger&vanderklis}. 
The differences between the two classes are probably driven by the mass accretion rate, which is close to the Eddington limit in Z sources, and relatively low in atoll sources \citep[e.g.][]{vanderklis1994,munoz-darias2014}. However, the distinction between Z and atoll is not clear-cut, because atoll sources also exhibit Z-shape tracks when observed on a long enough time-scale \citep{gierlinski&done2002,muno2002}. Furthermore, a single source may also transit between Z and atoll behaviour, as observed in XTE~J1701$-$462 \citep{lin2009,homan2010}. 

The X-ray spectra of NS-LMXBs 
in the soft state
are generally well described by two main components, a thermal one dominating in the soft band below $\sim 1$ keV, and a hard one possibly due to Comptonization of soft photons in a plasma with a temperature of $\sim 3$ keV \citep[e.g.][]{barret2001,paizis2006,farinelli2008}. In the so-called eastern model, the soft component is a multi-temperature black body produced by the accretion disc, while the hard component arises from Comptonization of NS seed photons \citep{mitsuda1984,mitsuda1989}. Physically, this scenario could correspond to Comptonization occurring in a boundary layer (BL) between the disk and the NS \citep{pringle1977,shakura&sunyaev1988,popham&sunyaev2001} or a more vertically extended spreading layer (SL) around the NS \citep{inogamov&sunyaev1999,suleimanov&poutanen2006}. Alternatively, the so-called western model assumes that the soft component is a single-temperature black body due to the NS, while the hard component originates from Comptonization of disk photons \citep{white1988}. However, more complex scenarios have been proposed. For example, \citet{lin2007} discuss a three-component model including both NS and disk thermal emission plus Comptonization, while \citet{cocchi2011} propose a model in which both NS and disk seed photons are Comptonized in an extended corona. Furthermore, on top of the primary continuum, reflection of X-ray photons off the accretion disk is detected in a number of sources \citep[e.g.][]{disalvo2009,miller2013,mondal2017,mondal2018,ludlam2017,ludlam2019,ludlam2022}, albeit not ubiquitously \citep{homan2018,ludlam2019}. 

The exact nature of the X-ray emitting regions in NS-LMXBs thus remains elusive, because the models outlined above are spectroscopically degenerate. X-ray polarimetry, on the other hand, can significantly constrain the geometry of the system, and thus its physical properties. X-ray polarimetric studies are now possible thanks to the \textit{Imaging X-ray Polarimetry Explorer} (\ixpe, \citealt{weisskopf2022,soffitta2021}) launched in December 2021. The first two non-pulsating NS-LMXBs observed by \ixpe\ have been GS~1826$-$238 \citep{capitanio2023} and Cyg~X-2 \citep{farinelli2023}, both observed in a high soft state. For GS~1826$-$238, \citet{capitanio2023} find an upper limit to the polarization of 1.3\% (at 99.73\% confidence level) over the \ixpe\ 2--8 keV energy range. This result is consistent with a quasi-spherical geometry of the X-ray source, such as an extended SL, or with a non-spherical source seen at a small viewing angle $(\lesssim 40\degr$). In the case of Cyg~X-2, \citet{farinelli2023} find a polarization degree of $1.8\% \pm 0.3$\%, with a polarization angle consistent with the direction of the radio jet, and thus most likely parallel to the rotation axis. This rules out the accretion disk itself as the main source of polarized X-rays, as well as a geometry in which the boundary layer is coplanar with the disk \citep{farinelli2023}, because in the optically thick case polarization is parallel to the disk \citep{chandra1960}.
Interestingly, \citet{long2022} find a similar result for Sco~X-1, based on \textit{PolarLight} data in the 3--8 keV band.

GX~9+9 is a bright atoll source \citep{hasinger&vanderklis}, whose light curve shows a 4.2 hr modulation in both the optical and X-ray bands \citep{hertz&wood1988,schaefer1990}. Its distance is not well known, however the estimates range between 5 kpc \citep{christian&swank1997} and 10 kpc \citep{savolainen2009}. This source has been consistently observed in a bright soft state \citep{gladstone2007,savolainen2009,iaria2020}. The average X-ray flux in the 2--20 keV band is $\sim$200\,mCrab \citep{iaria2020}, and the X-ray spectrum is well represented by a two-component emission model plus reflection \citep{kong2006,savolainen2009,iaria2020}. In fact \citet{iaria2020}, studying the broad-band X-ray spectrum with \textit{XMM-Newton} and \textit{BeppoSAX}, report the presence of a significant relativistic reflection component and estimate an inclination of 40\degr--50\degr, consistent with the upper limit of 70\degr\ indicated by the lack of X-ray eclipses \citep{schaefer1990,savolainen2009}. 
In this paper, we report on the spectral and polarimetric analysis of GX~9+9 with simultaneous \nustar\ and \ixpe\ observation. We also discuss numerical simulations specifically developed for the source. The \ixpe\ polarimetric data have been recently presented by \cite{chatterjee2023}, together with \textit{AstroSat} non-simultaneous spectral data taken in 2020. However, the \nustar\ capability to detect the reflection component and the simultaneity with \ixpe\ give us the possibility to relate the polarization properties to specific spectral components with good confidence. 

The paper is structured as follows. In Sect.~\ref{sec:data}, we describe the observation and the \ixpe\ and \nustar\ data reduction. In Sect.~\ref{sec:analysis}, we report on the analysis of the spectral and polarimetric data. Sect.~\ref{sec:disc} is devoted to the discussion of the results and the summary. 

\section{Observation and data reduction}
\label{sec:data}

\begin{table}
\caption{\label{tab:obs-log}
Logs of the \textit{IXPE} and \textit{NuSTAR} observations.
}
\begin{center}
  \begin{tabular}{cccc}
    \hline    \hline
 Satellite & Obs. Id. & Start time & Net exp. \\
 &&(UTC)&(ks)\\
\hline 
\textit{IXPE} & 01002401 & 2022-10-09T12:09:58 & 92.5 \\
\textit{NuSTAR} & 30801021002 & 2022-10-09T10:21:09 & 38.5 \\
\hline 
\end{tabular}
\end{center}
\end{table}

\subsection{\ixpe}
\ixpe\ observed the source on 2022 October 9 with its three Detector Units (DU)/Mirror Module Assemblies (MMA), for a net exposure time of 92.5 ks (Table \ref{tab:obs-log}). We produced cleaned level 2 event files using standard filtering criteria with the dedicated {\sc ftools} (v6.31) tasks\footnote{\url{https://heasarc.gsfc.nasa.gov/docs/ixpe/analysis/IXPE-SOC-DOC-009-UserGuide-Software.pdf}} and the latest calibration files (CALDB 20221021) and response matrices (v12). 
The $Q$ and $U$ Stokes spectra produced by the instrument pipeline are in the FITS format, and associated to an ancillary response matrix and a modulation response function (namely the product of the effective area and the modulation factor). The joint spectropolarimetric analysis of the Stokes parameters can thus be carried out with standard techniques and software tools.
We extracted the Stokes $I$, $Q$ and $U$ spectra from circular regions with a radius of 120\arcsec. We did not subtract the background, following the prescription by \cite{dimarco2023} for bright sources. However, we verified that background subtraction does not significantly alter the results, especially the polarimetric measurements. Given the source brightness, each energy bin of the flux ($I$) spectra contains more than 40 counts, ensuring the applicability of the $\chi^2$ statistics. Thus, we did not rebin the $I$ spectra, while we applied a constant energy binning of 0.2 keV for $Q$ and $U$ Stokes spectra. We fitted the $I$, $Q$ and $U$ Stokes spectra from the three DU/MMAs independently. 

\subsection{\nustar}
\nustar\ \citep{harrison2013} observed the source with its X-ray telescopes on Focal Plane Module A (FPMA) and B (FPMB), with a net exposure of 38.5 ks simultaneously with the first half of the \ixpe\ exposure. We produced cleaned event files with the dedicated \texttt{nupipeline} task and the latest calibration files (CALDB 20230208). In the case of \nustar\ the background is not negligible at all energies, therefore we performed background subtraction as follows. For both detectors, we extracted the background from a circular region with a standard radius of 1\farcm22, while we set the source radius at 2\arcmin\ following a procedure that maximizes the signal to noise ratio \citep{pico2004}. We rebinned the spectra with the standard task \texttt{ftgrouppha}, implementing the optimal scheme by \cite{kaastra&bleeker2016} with the additional requirement of a minimum signal to noise of 3 in each bin. The FPMA and FPMB spectra were fitted independently. We used the data in the 3--30 keV range, since the background starts dominating above 30 keV. 

\section{Data analysis}
\label{sec:analysis}

\begin{figure}
\includegraphics[width=1.0\columnwidth]{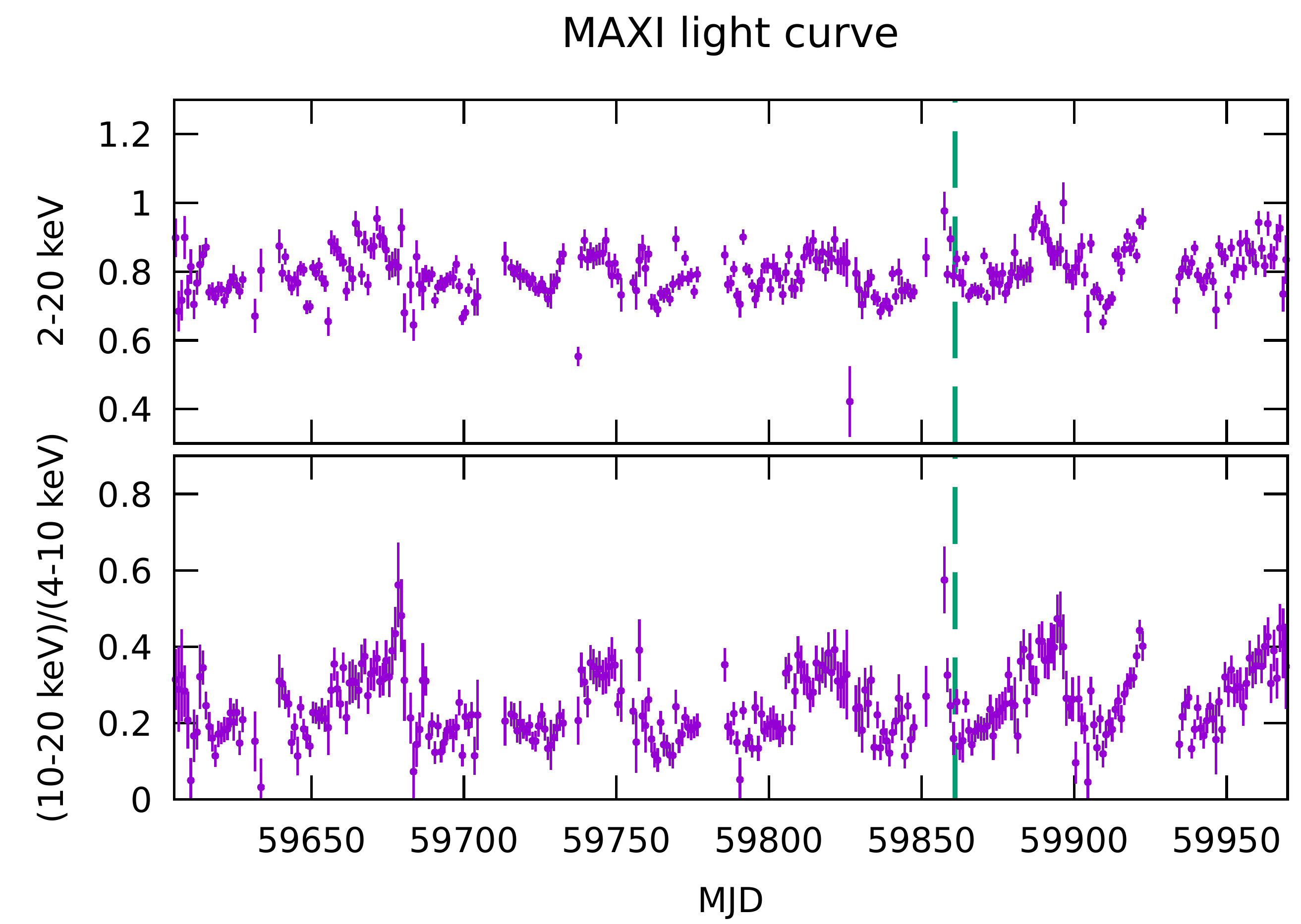}
\caption{
{MAXI} daily light curve of GX~9+9 (photon\percm \pers). 
\textit{Top panel:} 2--20 keV photon flux.
\textit{Bottom panel:} hardness ratio between the 10--20 keV and 4--10 keV photon fluxes. The vertical dashed line marks the date of the \ixpe\ and \nustar\ observation.
}
\label{fig:maxi_lc}
\end{figure}

\begin{figure}
\includegraphics[width=1.0\columnwidth]{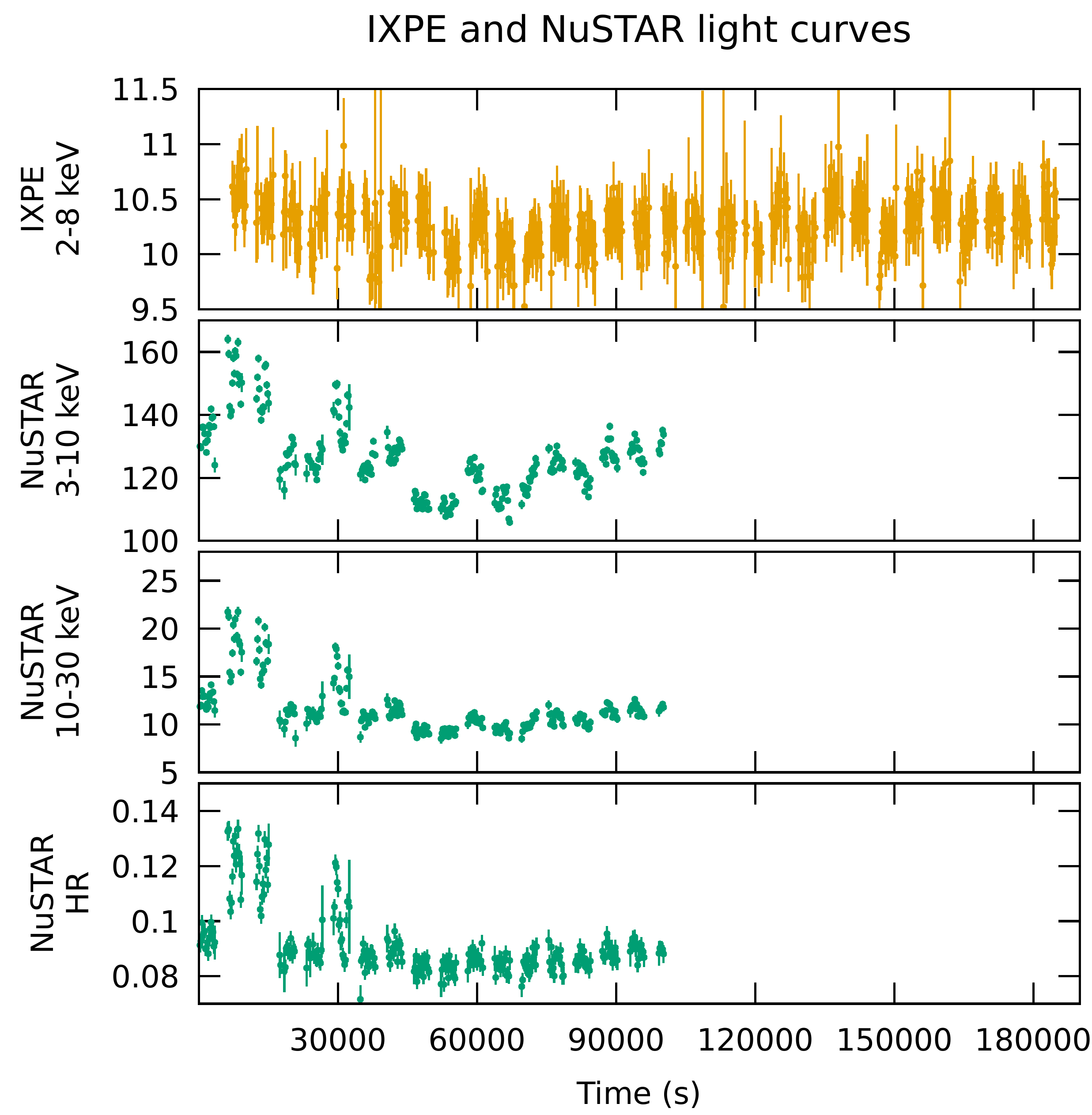}
\caption{
\ixpe\ and \nustar\ light curves of GX~9+9 (count\pers). The lower panel shows the \nustar\ hardness ratio (10--30 keV)/(3--10 keV).
 Time bins of 200 s are used. 
}
\label{fig:nustar_lc}
\end{figure}

\begin{figure}
\includegraphics[width=1.0\columnwidth]{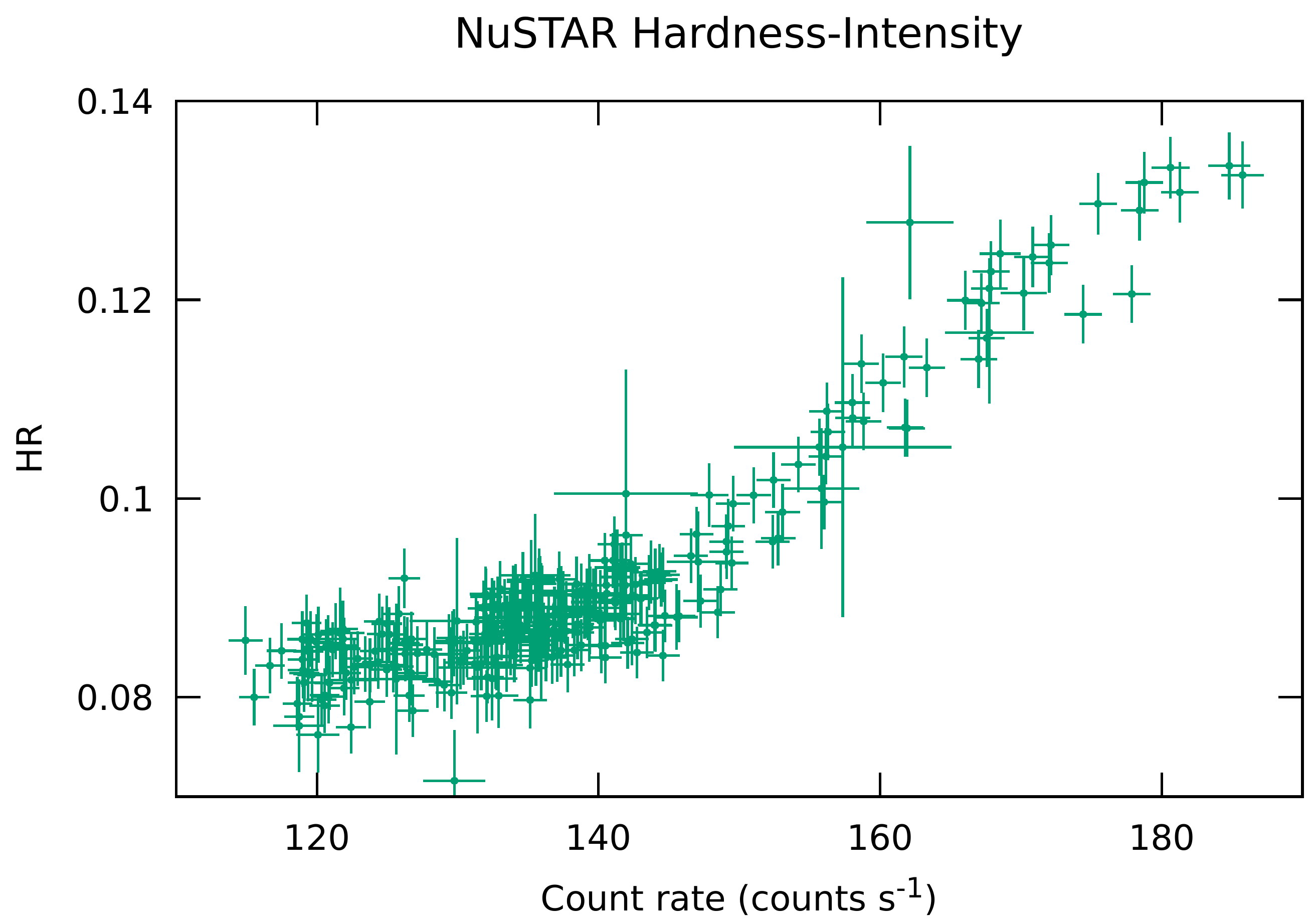}
\caption{
\nustar\ count rate hardness ratio (10--30 keV/3--10 keV) versus total count rate (3--30 keV).
Time bins of 200 s are used. 
}
\label{fig:nustar_hid}
\end{figure}

\begin{figure}
\includegraphics[width=1.0\columnwidth]{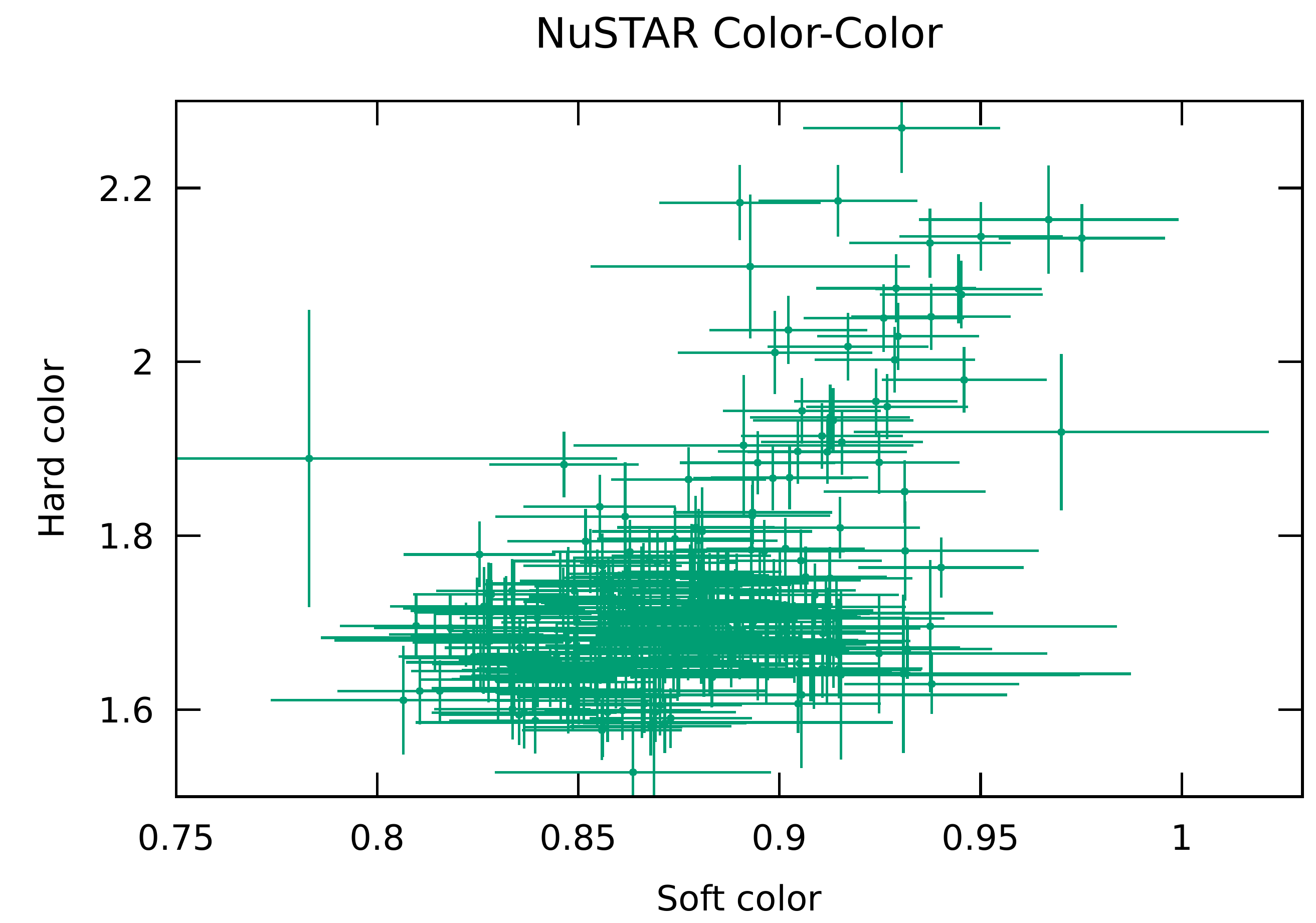}
\caption{
\nustar\ hard colour (6--30 keV/4.5--6 keV) versus soft colour (4.5--6 keV/3--4.5 keV).  
Time bins of 200 s are used. 
}
\label{fig:nustar_cc}
\end{figure}

GX~9+9 is known to exhibit complex long-term X-ray variability \citep{kotze&charles2010}, however as of 2009 the light curve of the 
{Monitor of All-sky X-ray Image} \cite[{MAXI},][]{matsuoka2009} shows a quite constant baseline and small amplitude variations \citep{asai2022}. The {MAXI} light curve spanning one year between 2022 Jan 26 and 2023 Jan 26 is shown in Fig. \ref{fig:maxi_lc}. The X-ray flux is quite stable, while the hardness ratio (HR) shows variations by a factor of 2 on time-scales of roughly 1--2 months. 
In Fig. \ref{fig:nustar_lc} we show the \nustar\ light curve of the source, together with the HR. The \nustar\ hardness-intensity diagram and the colour-colour diagram are shown in Figs. \ref{fig:nustar_hid} and \ref{fig:nustar_cc}, respectively. Some spectral variability is apparent for a short interval at the beginning of the \nustar\ exposure. However, the variation in HR is not dramatic, and no flaring is detected. Therefore, in the following analysis we consider the  spectrum averaged over the whole observation.

\begin{figure}
\includegraphics[width=1\columnwidth, trim={0 1cm 0 0}]{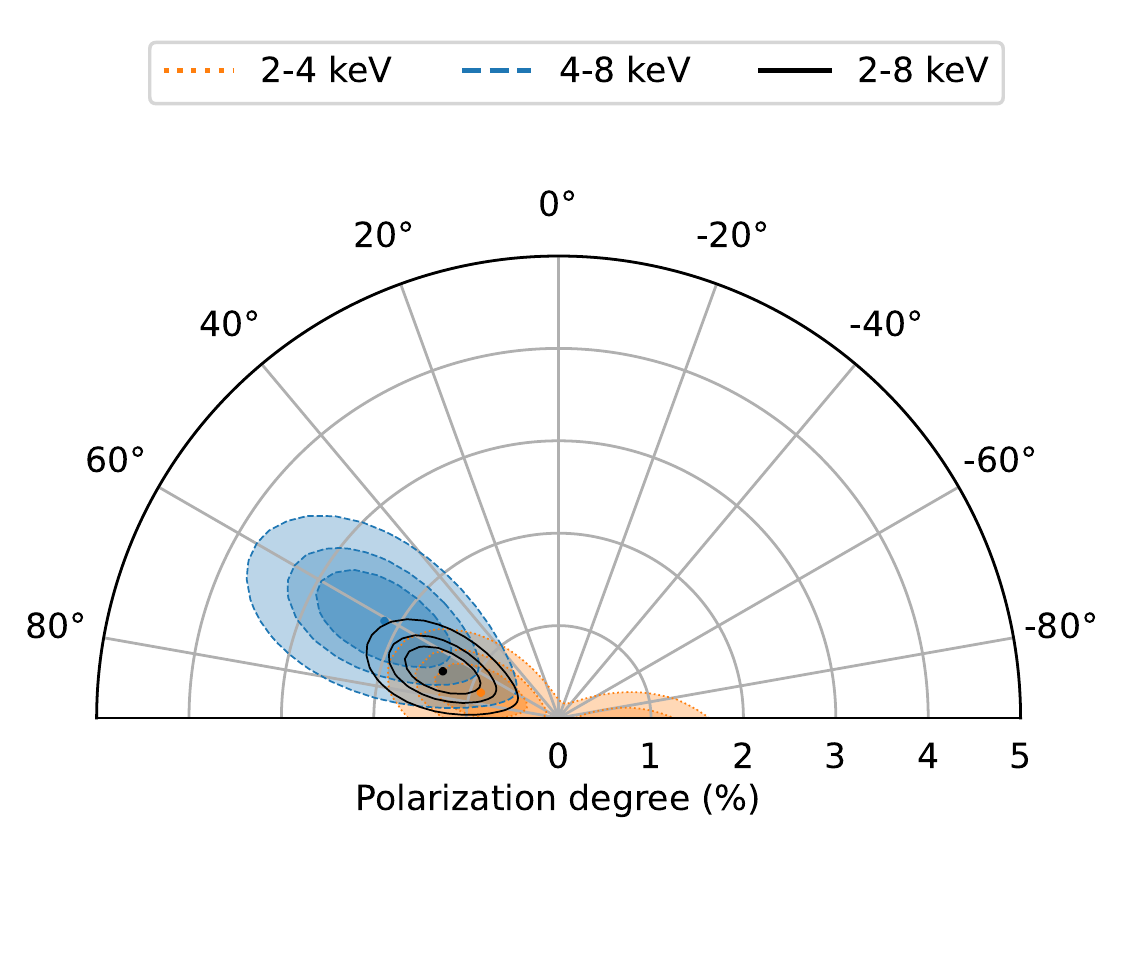}
\caption{
Contour plots of the PD and PA at the 68, 90 and 99\% confidence levels, in the 2--4 keV (orange dotted), 4--8 keV (blue dashed) and 2--8 keV (black solid) energy bands.
\label{fig:conts_PD_PA}
}
\end{figure}

In Table \ref{tab:poldeg-ang}, we report the polarization degree (PD) and polarization angle (PA), as measured from \ixpe\ spectra using \xspec\ 12.13.0 \citep{arnaud1996}, with the 68\% confidence level uncertainty for one parameter of interest. 
These values are well consistent within the errors with those found from the polarization cubes extracted with \textsc{ixpeobssim} \citep{ixpeobssim}. 
The two-dimensional contour plots of PD and PA are shown in Fig. \ref{fig:conts_PD_PA}. 
The results are consistent within the errors with those reported by \citet{chatterjee2023}.

\begin{table}
\caption{\label{tab:poldeg-ang}
PD and PA measured with \textsc{xspec}.}
\begin{center}
  \begin{tabular}{ccc}
    \hline    \hline
 Energy range (keV)   & PD (\%)   & PA (deg) \\
 \hline 
2--8 &  $1.4 \pm 0.3$ &  $68 \pm 6$  \\
2--4  &$<1.8$ & -- \\
4--8  &$2.2 \pm 0.5$ & $61 \pm 7$  \\
\hline 
\end{tabular}
\end{center}
\tablefoot{
The uncertainties are given at 68\% ($1 \sigma$) confidence level for one parameter of interest.
The upper limit is quoted at 99\% confidence level.
}
\end{table}

\subsection{The \nustar\ spectrum}

\begin{figure}
\includegraphics[width=1.0\columnwidth]{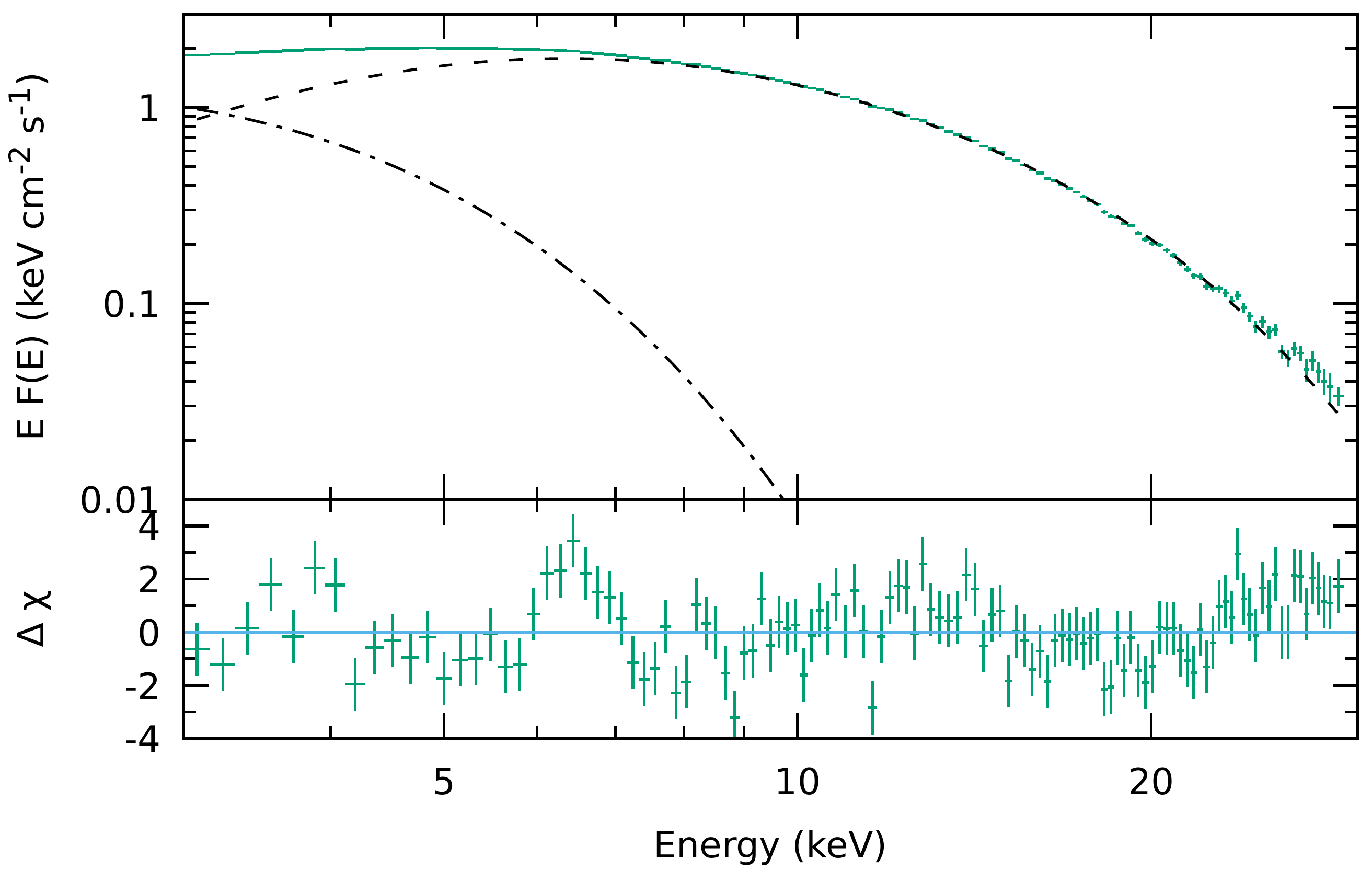}
\caption{
\textit{Top panel:} Deconvolved \nustar\ spectrum fitted with a model composed of \diskbb\ (dash dotted line) and \comptt\ (dashed line).
\textit{Bottom panel:} Residuals in units of $\sigma$. The data from the two detectors FPMA and FPMB are combined (with \texttt{setplot group} in \xspec) for plotting purposes only.}
\label{fig:nustar_norefl_eeuf}
\end{figure}

As a first step, we assess the spectral properties of the source by fitting the \nustar\ data in the 3--30 keV band. In all of our models we include interstellar absorption \citep[\tbabs\ model in \xspec;][]{tbabs}, with a column density fixed at $3 \times 10^{21}$ \percm\ \citep{iaria2020} because the energy band does not extend to low energies where absorption is strong, thus the fit is not very sensitive to this parameter. We start by fitting the \nustar\ spectrum with a standard two-component model composed by a disk multicolour black body \citep[\diskbb\ in \xspec;][]{mitsuda1984} and a Comptonized spectrum \citep[\comptt\ in \xspec;][]{titarchuk1994}, including a cross-calibration constant between FPMA and FPMB which is found to be $1.006\pm 0.001$. However, we find a discrepancy between the FPMA and FPMB spectra at low energies (3--4 keV), the FPMA flux being significantly larger. This could be an effect of a known rip in the multi-layer insulation that encloses the optics \citep{madsen2020}, not fully accounted for by the current calibration. Following \cite{madsen2020}, we include in our model the multiplicative table \texttt{numliv1} specifically designed to correct this issue.\footnote{\url{https://nustarsoc.caltech.edu/NuSTAR_Public/NuSTAROperationSite/mli.php}} 

The simple model \texttt{diskbb+comptt} does not provide a good fit to the \nustar\ data, because it leaves clear residuals especially in the  6--7 keV range, giving a total $\chi^2$/d.o.f. = 319/228 (see Fig. \ref{fig:nustar_norefl_eeuf}).
We note that in \comptt\ the seed photons are distributed according to Wien's law, so that the model low energy tail goes as $E^3$. To check if 
the observed residuals may be due to the interplay with the softer disk emission,
we replaced \comptt\ with \comptb\ \citep{farinelli2008}, in which the seed photon spectrum is a blackbody. However, the behaviour of the residuals  did not change substantially.
It is also worth noting that \citet{iaria2020} find that a relativistic smeared reflection component is needed to fit the broadband spectrum, but they do not find clear evidence of a Fe K emission line, while this is apparent in the \nustar\ spectrum.
Including a narrow Gaussian line at 6.4 keV improves the fit, which however remains not very good ($\chi^2$/d.o.f. = 282/226). 

A further improvement is found replacing the narrow Gaussian with a relativistically broadened emission line \cite[\relline,][]{relline}. 
In \texttt{relline} we fix the emissivity index at 3 and the dimensionless spin at 0.2, because the fit is not very sensitive to these parameters. 
We notice that the line energy and the disk inclination are degenerate in the spectral fit, thus we initially fix the inclination at 40\degr.
We find a fit with $\chi^2$/d.o.f. = 261/225; the inner disk radius of \relline\ is $6 \pm 3$ in units of the innermost stable circular orbit (ISCO), the line energy is $6.51^{+0.09}_{-0.10}$ keV and its equivalent width is $16 \pm 3$ eV. The line energy might be consistent with an origin from ionized material. Indeed, if we assume an inclination of 30\degr, we obtain a statistically equivalent fit with a line energy of $6.70^{+0.05}_{-0.06}$ keV, i.e. consistent with the Fe~XXV K$\alpha$ line. However, if we assume an inclination greater than 50\degr, the line is consistent with being neutral. These results motivate us to test a more self-consistent reflection model to try to break the degeneracy.
We also check whether the emission line could be actually an artefact of absorption, but replacing the emission line with an absorption edge at $\sim 7$ keV results in a worse fit ($\chi^2$/d.o.f. = 274/226) with significant residuals in the 6--7 keV band. 

\begin{table}
\caption{\label{tab:Fit}
Best-fitting model parameters of the fits to the \nustar\ and \nustar+\ixpe\ data.}
\begin{center}
  \begin{tabular}{lll}
\hline
\hline
Parameter & \nustar & \nustar+\ixpe \\ 
\hline 
   \multicolumn{3}{c}{\diskbb} \\
$kT_{\rm in}$ (keV) & $1.00\pm 0.04$ & $1.020\pm 0.005$ \\
$N_{\rm d}$ & $190\pm 25$ & $168\pm 5$ \\
\hline
 \multicolumn{3}{c}{\comptb} \\
 $kT_{\rm s}$ (keV) & $1.59 \pm 0.07$ & $1.575 \pm 0.005$ \\
 $\alpha$ & $2.5 \pm 0.3$ & [2.5] \\
 $kT_{\rm e}$ (keV) & $3.4 \pm 0.3$ & [3.4] \\
 $N_{\rm c}$ ($10^{-2}$) & $3.7 \pm 0.1$ & $3.77 \pm 0.03$ \\
 \hline
\multicolumn{3}{c}{\relxillns} \\
$q_{\rm em}$ & [3] & [3] \\
$a$ & [0.2] & [0.2]\\
incl (deg) & $28 \pm 8$ & [28]\\
$R_{\rm in}$ (units of $R_{\rm ISCO}$)  & $5 \pm 3$ & [5]\\
$kT_{\rm bb}$ (keV) & =1.59 & =1.575\\
$\log (\xi$/erg cm \pers) & $3.3 \pm 0.3$ & [3.3]\\
$A_{\rm Fe}$ & [1] & [1] \\
$\log n_{\rm e}$ & [18] & [18] \\
 $N_{\rm r}$ ($10^{-4}$) & $4 \pm 1$ & $3.3 \pm 0.3$ \\
 \hline
\multicolumn{3}{c}{Cross-calibration} \\
\texttt{numliv1} & $0.94 \pm 0.01$ & [0.94]\\
$C_{\rm FPMB-FPMA}$ & $1.006 \pm 0.001$ & [1.006]\\
$C_{\rm DU1-FPMA}$ && $0.907\pm 0.005$ \\
$C_{\rm DU2-FPMA}$ && $0.877\pm 0.005$ \\
$C_{\rm DU3-FPMA}$ && $0.827 \pm 0.004$ \\
\multicolumn{3}{c}{Gain shift} \\
$\alpha_{\rm DU1}$ && $0.989^{+0.001}_{-0.002}$ \\
$\beta_{\rm DU1}$ (eV) && $-24^{+7}_{-8}$ \\
$\alpha_{\rm DU2}$ && $0.987\pm 0.002$ \\
$\beta_{\rm DU2}$ (eV) && $-16\pm 8$ \\
$\alpha_{\rm DU1}$ && $0.990^{+0.002}_{-0.001}$ \\
$\beta_{\rm DU1}$ (eV) && $-15^{+8}_{-6}$ \\
\hline
$\chi^2$/d.o.f. & $229/224$ & 865/841\\
\hline
\multicolumn{3}{c}{Photon flux ratios\tablefootmark{a}} \\
\multicolumn{3}{c}{2--8 keV} \\
$F_{\rm diskbb}/F_{\rm tot}$ & 0.51& \\
$F_{\rm comptb}/F_{\rm tot}$ & 0.43& \\
$F_{\rm relxillns}/F_{\rm tot}$ & 0.06& \\
\multicolumn{3}{c}{2--4 keV} \\
$F_{\rm diskbb}/F_{\rm tot}$ & 0.66& \\
$F_{\rm comptb}/F_{\rm tot}$ & 0.30& \\
$F_{\rm relxillns}/F_{\rm tot}$ & 0.04& \\
\multicolumn{3}{c}{4--8 keV} \\
$F_{\rm diskbb}/F_{\rm tot}$ & 0.26& \\
$F_{\rm comptb}/F_{\rm tot}$ & 0.66& \\
$F_{\rm relxillns}/F_{\rm tot}$ & 0.08& \\
\multicolumn{3}{c}{Energy flux (2--8 keV)} \\
$F_{\rm tot}$ (\fluxcgs) & $4.1\times 10^{-9}$& \\
\hline 
\end{tabular}
\tablefoot{Uncertainties are given at 68\% confidence level for one parameter of interest. Parameters in square bracket are kept frozen during the fit.\\
\tablefoottext{a}{The photon fluxes are in units of photon\percm\pers.}
}
\end{center}
\end{table}

\begin{figure*}
\includegraphics[width=1\columnwidth,trim={0 -1.8in 0 0},clip]{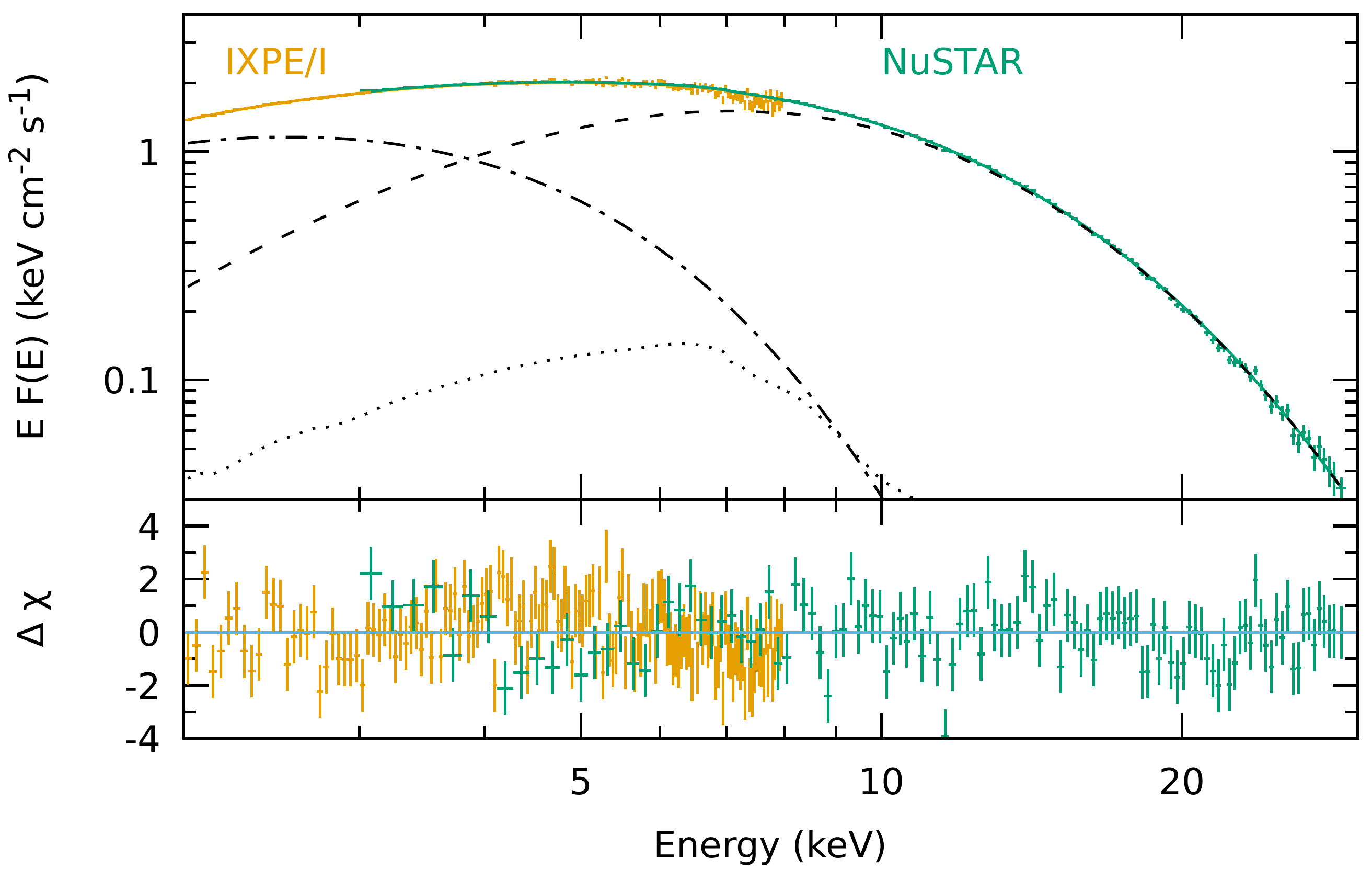}
\includegraphics[width=1\columnwidth]{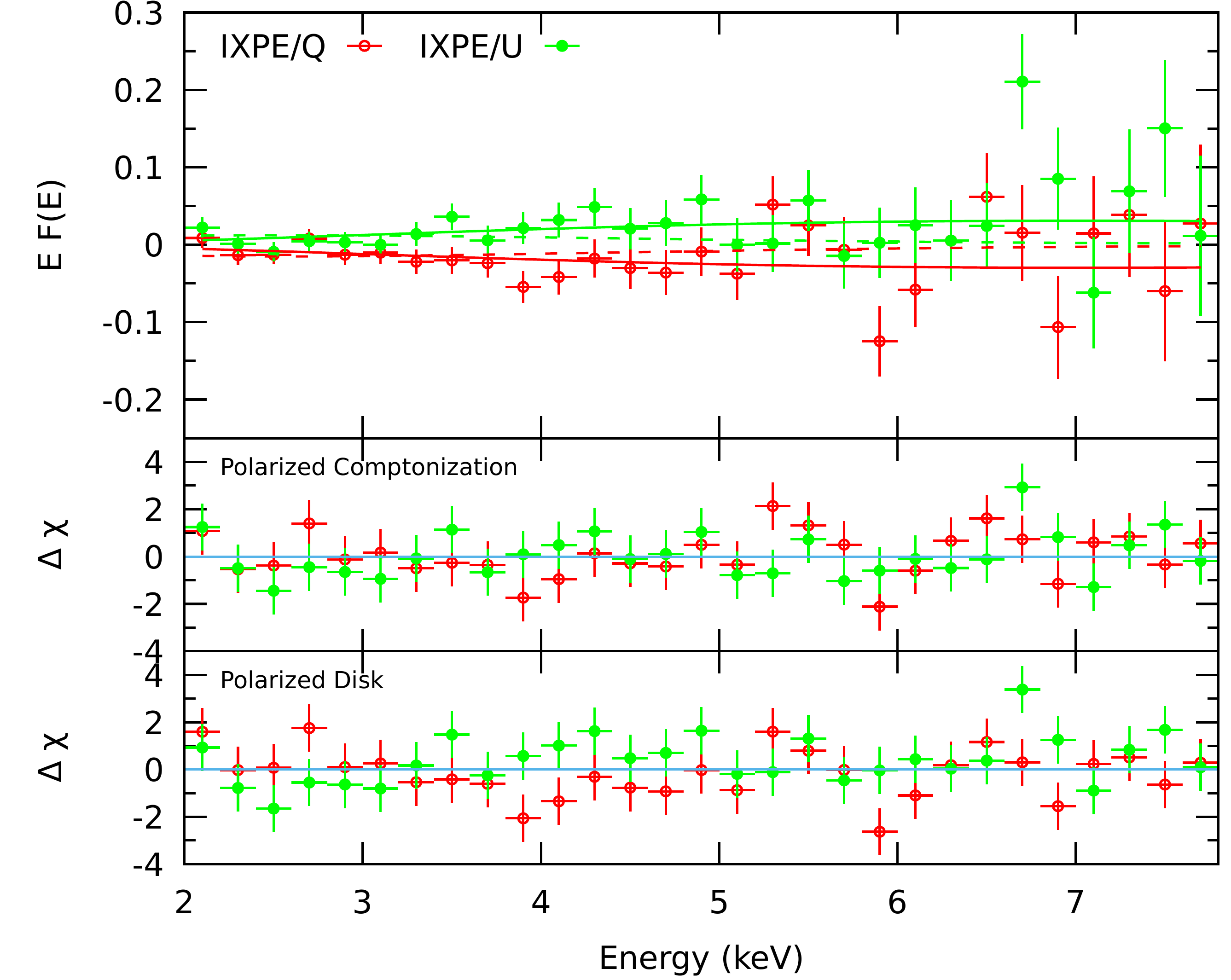}
\caption{
\textit{Left panel:} \ixpe\  and \nustar\ deconvolved spectra with best-fitting model (see Table \ref{tab:Fit}) and residuals. The model includes \diskbb\ (dashed dotted line), \comptb\ (dashed line) and \relxillns\ (dotted line). 
\textit{Right panel:} \ixpe\ $Q$ and $U$ Stokes spectra with the best-fitting model, assuming as the only polarized component either \comptb\ (solid lines) or \diskbb\ (dashed lines), and corresponding residuals. Data from different detectors are combined (with \texttt{setplot group}) for plotting purposes.
}
\label{fig:nustar+ixpe_spectra}
\end{figure*}

Finally, we replace the line with a complete reflection model. The residuals observed in Fig. \ref{fig:nustar_norefl_eeuf} do not show the presence of an obvious Compton hump, indicating that reflection is produced by a softer illuminating spectrum than the typical non-thermal (power-law) continuum assumed in standard reflection models. We thus employ \relxillns\ \citep{relxillns}, a flavor of \texttt{relxill} \citep{relxill_1,relxill_2} in which the primary continuum illuminating the disk is a single temperature blackbody spectrum (physically corresponding to the NS surface or a SL). 
The code assumes an illuminating spectrum incident at 45\degr\ on the surface of the disk \citep{relxillns}. In \relxillns\ we fix some parameters, which the fit does not constrain, at reasonable values. We set the emissivity index $q_{\rm em}=3$ \citep[see, e.g.,][]{wilkins2018}, the dimensionless spin $a=0.2$, the iron abundance $A_{\rm Fe}=1$, and the number density $\log n_{\rm e}=18$ \citep[e.g.][]{ludlam2022}\footnote{
The adopted number density is consistent with the inner region of a standard disk \citep[see][and references therein]{relxillD}. Our fit is not sensitive to this parameter, because its effects on the X-ray reflection spectrum are only significant at lower energies \citep{ballantyne2004,relxillD}. 
}. We also tie the blackbody temperature to that of the seed photon spectrum of \comptb.
We leave free to vary the inclination, the inner disk radius 
(in units of the ISCO)
and the ionization parameter. The inclusion of the \relxillns\ component provides an excellent fit, with $\chi^2$/d.o.f. = 229/224. We report the best-fitting parameters in Table~\ref{tab:Fit}. From the spectral index $\alpha$ and electron temperature of \comptb, using equations (17) and (22) of \cite{tl1995}, we obtain an optical depth $\tau$ equal to $2.1\pm0.3$ and $5.1\pm0.4$ for slab and spherical geometry, respectively.

The best-fitting parameters are quite typical for a NS-LMXB in the soft state. Assuming a distance to the source of 7.5 kpc, the normalization of \diskbb\ corresponds to an inner disk radius of $\sim 10 \sqrt{\cos i}$ km (where $i$ is the inclination). 
As for the Comptonization component, we can estimate the size of the seed photon-emitting region from the best-fit parameters of \comptb. Assuming that all seed photons are Comptonized, we compute the flux of the seed black body spectrum, hence the luminosity and the emission area \citep[see also][]{intzand1999}. We estimate that the seed photon-emitting region has an equivalent spherical radius of 5 km. This is consistent with the seed photons originating in a region smaller than the entire NS, such as the boundary layer.

The parameters that we derive with \nustar\ are not the same as those found by \citet{chatterjee2023},
who fitted \textit{AstroSAT} data with a two-component model consisting of \diskbb\ plus the Comptonization model \texttt{nthcomp}. 
For example, we find an inner disk temperature of $1.00\pm0.04$ keV, while \citet{chatterjee2023} report $0.65\pm0.04$ keV. Concerning the Comptonization component, we find a seed photon temperature $kT_{\textrm{s}}=1.59\pm0.07$ keV, an electron temperature $kT_{\textrm{e}}=3.4\pm0.3$ keV and an optical depth $\tau=5.1\pm0.4$ (assuming spherical geometry); \citet{chatterjee2023} find  $kT_{\textrm{s}}=1.05\pm0.03$ keV, $kT_{\textrm{e}}=5.58\pm0.82$ keV and $\tau=3.72\pm0.69$ (also in spherical geometry). We note, however, that the \textit{AstroSAT} data used by \citet{chatterjee2023} were taken in July 2020, namely more than two years before the \ixpe\ and \nustar\ observations. \citet{chatterjee2023} perform the spectro-polarimetric fit using the time-average spectrum, however they also report a time-resolved analysis of the \textit{AstroSAT} spectrum splitted in four time intervals. From their Table 1, the spectral parameters during the fourth interval (i.e. the hardest spectral state) are most similar to the parameters we derive with \nustar.
It is thus likely that the differences between our results and those of \citet{chatterjee2023} are due to the spectral variability of the source. 

\subsection{Spectropolarimetric analysis}\label{sec:spectropol}

Once we obtained a satisfactory baseline model (\diskbb+\comptb+\relxillns) for the source spectrum with \nustar\ data,
we used it to jointly fit the \nustar\ and \ixpe\ spectropolarimetric data. 
As any of the three emission components can be polarized, we first employ the following  \xspec\ model:
\begin{align*}
    \texttt{c\_cal} \times \tbabs \times (\polconst^{(d)} \times \diskbb \\ + \polconst^{(c)} \times \comptb + \polconst^{(r)} \times \relxillns) , 
\end{align*}
where \texttt{c\_cal} denotes the cross-calibration constants. We note that \relxillns\ includes the fluorescent iron line, which should be unpolarized; however, the line contribution to the photon flux of \relxillns\ is only $\sim 10\%$ in the 6--7 keV band.

We find significant residuals at low and high energies in the \ixpe/$I$ spectra, which very likely are an artifact of calibration issues already observed in other sources \citep{taverna2022,krawczynski2022,marinucci2022}. To correct them, we apply a gain shift to the response files of \ixpe/$I$ spectra with the \texttt{gain fit} command in \xspec, and link the gain parameters of $Q$ and $U$ spectra to those of the $I$ spectra. The energy shift is calculated using the relation $E' = E/\alpha - \beta$, where in our case $\alpha$ is not far from 1 and $\beta$ has an absolute value of 15--20 eV (see Table \ref{tab:Fit}).

In the model above, \polconst\ assumes that PD and PA of each component are energy-independent. To reduce degeneracy effects in the narrower \ixpe\ band, we keep some parameters of \relxillns\ and \comptb\ fixed at their best value obtained with \nustar\ (see Table \ref{tab:Fit}). The overlap in 3--8 keV  range between \ixpe\ and \nustar\ spectra, as well
as 50\% of their exposures (see Fig. \ref{fig:nustar_lc}) is expected to minimize systematics in the derived
polarimetric \xspec\ parameters.
However, the \ixpe\ bandpass does not allow us to obtain tight constraints on PD and PA of each component,
so a series of assumptions based on theoretical and observational expectations is needed.

\begin{figure*}
\includegraphics[width=1\columnwidth]{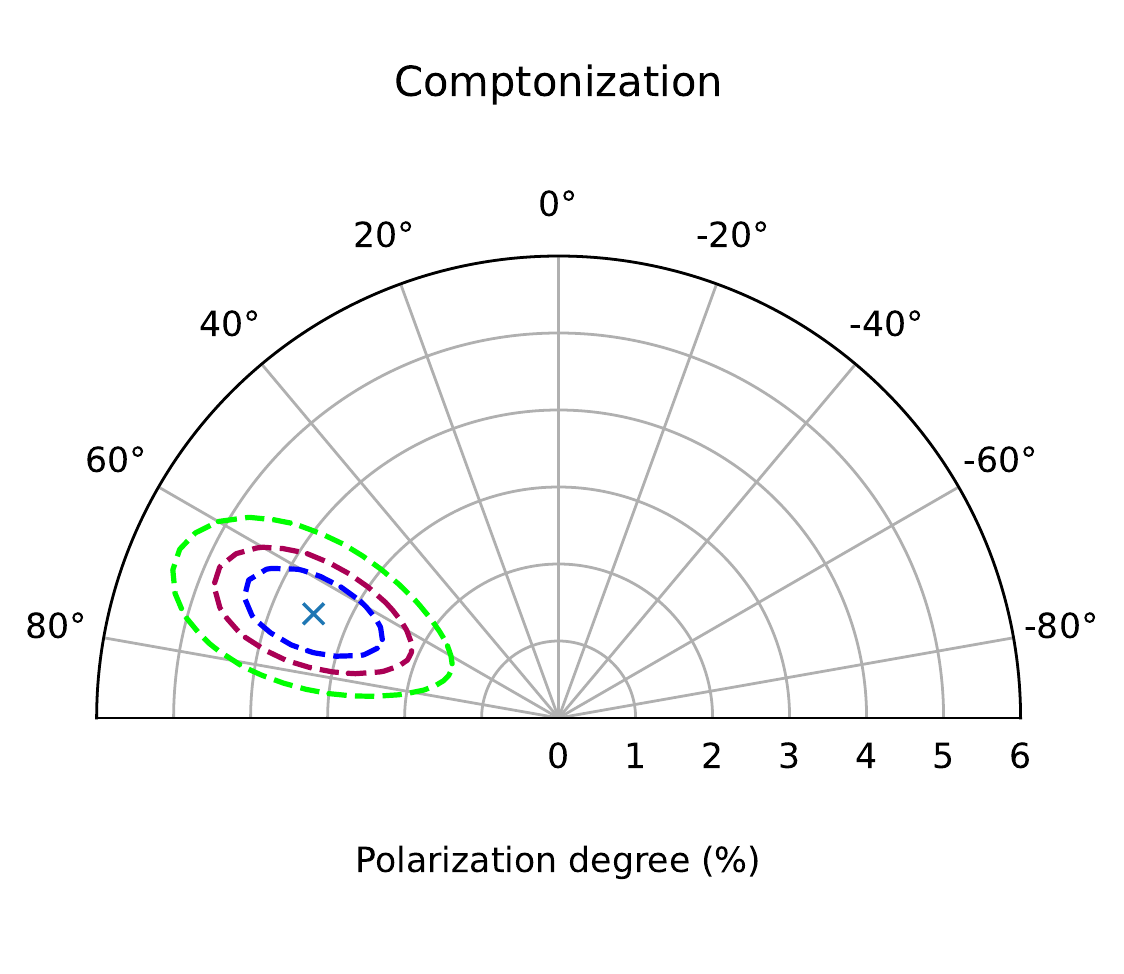}
\includegraphics[width=1\columnwidth]{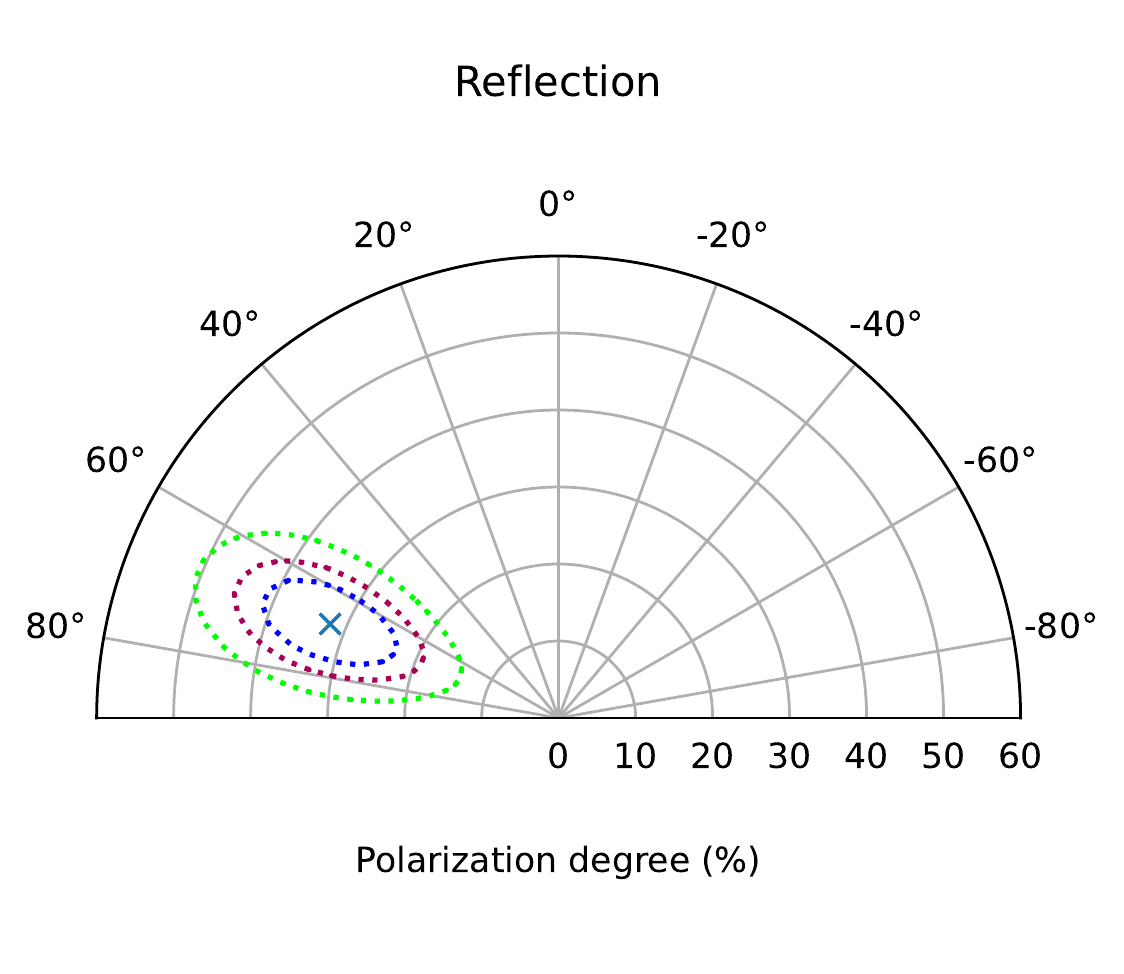}
\caption{
\textit{Left panel:} Contour plots of the PD and PA at 68, 90 and 99\% confidence levels for the Comptonization component, assuming that it is the only one polarized.
\textit{Right panel:} Similar contour plots for the reflection component.
Note the very different scale in the two plots.
}
\label{fig:cont_compt_refl}
\end{figure*}

As a first test,  we assume that only one of the three spectral components is polarized, the other two having null PD.  
Assuming first a polarized disk only, the fit provides $\chi^2$/d.o.f. = 879/841 
(151/96 for the subset of Stokes $Q$ and $U$ spectra).
The fit is acceptable, but the assumption is not consistent with the observed PD increasing with energy (see Fig. \ref{fig:conts_PD_PA}) -- actually, the opposite is
expected from the disk contribution to the total photon flux (see Table \ref{tab:Fit}).
Considering, on the other hand, the case where either \comptb\ or \relxillns\ are the only polarized component, we obtain a better fit with $\chi^2$/d.o.f. = 865/841
(137/96 for the Stokes $Q$ and $U$ spectra)
in both cases. We show in Fig. \ref{fig:nustar+ixpe_spectra} the best fit assuming \comptb\ as the only polarized component, noting that the residuals are essentially the same as the case where \relxillns\ is the only polarized component. 
The model provides a good fit of both the flux spectra (left panel of Fig. \ref{fig:nustar+ixpe_spectra}) and the Stokes $Q$ and $U$ spectra (right panel of Fig. \ref{fig:nustar+ixpe_spectra}). In particular, we do not find strong residuals for the $Q$ and $U$ spectra (right, center panel). For comparison, in Fig. \ref{fig:nustar+ixpe_spectra} we show also the residuals for the $Q$ and $U$ spectra assuming that \diskbb\ is the only polarized component (right, bottom panel). 
We show in Fig. \ref{fig:cont_compt_refl} the contour plots of the PD and PA in the two cases of Comptonization-only and reflection-only polarization. The reflection-dominated scenario requires a very large PD, because the reflection components is significant but  subdominant, contributing to roughly 10\% of the 4--8 keV flux. 

Beside the extreme cases tested above, there are many possible combinations that could explain the observed polarization. If we remove all assumptions and we leave free to vary all the \polconst\ components, the PD of the Comptonized and reflected radiation are not well constrained, because these two components are quite degenerate:
they peak at the same energy and have a similar spectral shape. Their total flux is clearly different, but their relative contribution to the polarized flux is uncertain. 

Then, because of the similarity in spectral shape of GX~9+9 to Cyg~X-2 (and in general to bright LMXBs in soft state), 
following \cite{farinelli2023}, and the fact that the reflection component has a PA perpendicular to the disk surface
\citep{matt1993,poutanen1996,schnittman&krolik2009}
we assume that the PA of the reflected and Comptonized radiation are the same. We note that this configuration is consistent with a geometry in which the BL/SL has a vertical height significantly greater than its radial extension, with $H\gg\Delta R$.
We start by fixing the PD of the Comptonized component at 1\%, assuming for the disk PD a conservative value of 1\% at the inferred
source inclination.
We obtain a PD of the reflection component of $30\% \pm 8\%$, while the PA is $63\degr \pm 7\degr$  for the BL+reflection and $-41\degr \pm 25\degr$ for the disk (68\% confidence). The disk PA is not tightly constrained, however it is consistent with being perpendicular to the observed PA. 
We then assume that the PA of the disk and BL+reflection components are perpendicular to each other. 
We now leave free to vary the PD of the disc, finding an upper limit of 3.6\% (99\% confidence on a single parameter) and a PD of the reflection component of $33\% \pm 12\%$ (68\% confidence). Then, we fix the PD of the Comptonized component at zero, obtaining an upper limit to the disk PD of 3.7\% and a PD of the reflection component of $44\% \pm 12\%$. Finally, we leave free to vary the PD of both the disk and Comptonized components, and we assume that the reflected radiation has a PD of 10\% \citep[see, e.g., Fig. 7 in][]{matt1993}. In this case, we obtain an upper limit to the disk PD of 2.9\% (99\% confidence) and a PD of the Comptonized component of $3\% \pm 1 \%$ (68\% confidence). 
In Table \ref{tab:PD-PA} we summarize the different tests discussed above.

\begin{table}
\caption{\label{tab:PD-PA}
Polarization degree and angle of each spectral component, for different scenarios described in Sect. \ref{sec:spectropol}.
}
\begin{center}
  \begin{tabular}{lll}
\hline
\hline
Component & PD (\%) & PA (deg) \\ 
\hline 
\diskbb & [1] & $-41 \pm 25$\\
\comptb & [1] & $63 \pm 7$ \\
\relxillns & $30 \pm 8$ & =PA$_{\comptb}$ \\
\hline
\diskbb & $<3.6$ & =PA$_{\comptb} - 90$\\
\comptb & [1] & $67 \pm 6$ \\
\relxillns & $33 \pm 12$ & =PA$_{\comptb}$ \\
\hline
\diskbb & $<3.7$ & =PA$_{\relxillns} - 90$\\
\comptb & [0] & -- \\
\relxillns & $44 \pm 12$ & $67 \pm 6$ \\
\hline 
\diskbb & $<2.9$ & =PA$_{\comptb} - 90$\\
\comptb & $3 \pm 1$ & $67^{+5}_{-6}$ \\
\relxillns & [10] & =PA$_{\comptb}$ \\
\hline
\end{tabular}
\tablefoot{Parameters in square bracket are frozen.
}
\end{center}
\end{table}

\section{Discussion and conclusions}
\label{sec:disc}

We investigated the X-ray spectro-polarimetric properties of the bright atoll NS-LMXB GX~9+9 using \ixpe\ and \nustar\ data.
The polarimetric measurements are in agreement with those reported by \cite{chatterjee2023}, however these authors fit non-simultaneous \textit{AstroSat} data with a two-component model that does not include reflection. The \nustar\ spectrum shows the presence of Fe K$\alpha$ emission line, and a reflection component is needed to properly fit the data \citep[see also][]{iaria2020}. 
To the best of our knowledge, this is the first detection of a relativistically broadened iron line in this source. The reflection component
is likely to play a significant role, as we argue in the following.

Like for the case of Cyg~X-2 \citep[][]{farinelli2023}, we only find an upper limit of 3--4\% to the disk PD,
which, because of the source inclination, does not allow us to eventually derive tight constraints about the properties
of a scattering atmosphere above the surface where the quasi-thermal spectrum is emitted \citep{ss73, pt74}.
The marginal hint of a disk PA at right angle with respect to the reflection component could anyway be indicative of the
presence of a scattering medium with $\tau \ga 2$ \citep{st85}.

On the other hand, the data are consistent with the Comptonized component being significantly polarized, up to 3--4\% depending on the assumptions. \citet{gnarini2022} and \citet{capitanio2023} discuss detailed numerical simulations of the X-ray polarization due to the Comptonizing region in NS-LMXB, assuming different geometrical configurations. In our case, the PA does not carry information because the orientation of the source is unknown, however the measured PD constrains the possible geometries. The PD depends on both the shape of the Comptonizing region and on the inclination of the source to the line of sight. 
For the latter parameter, the fit with the \relxillns\ model provides values in the range $i \approx 20\degr-40\degr$,
 namely an upper bound consistent with the value $i=40\degr$ reported by \cite{iaria2020}.
 In any case, the inclination is known to be less than 70\degr\ in this source \citep{schaefer1990,savolainen2009}. 
 
\begin{figure}
\includegraphics[width=1\columnwidth]{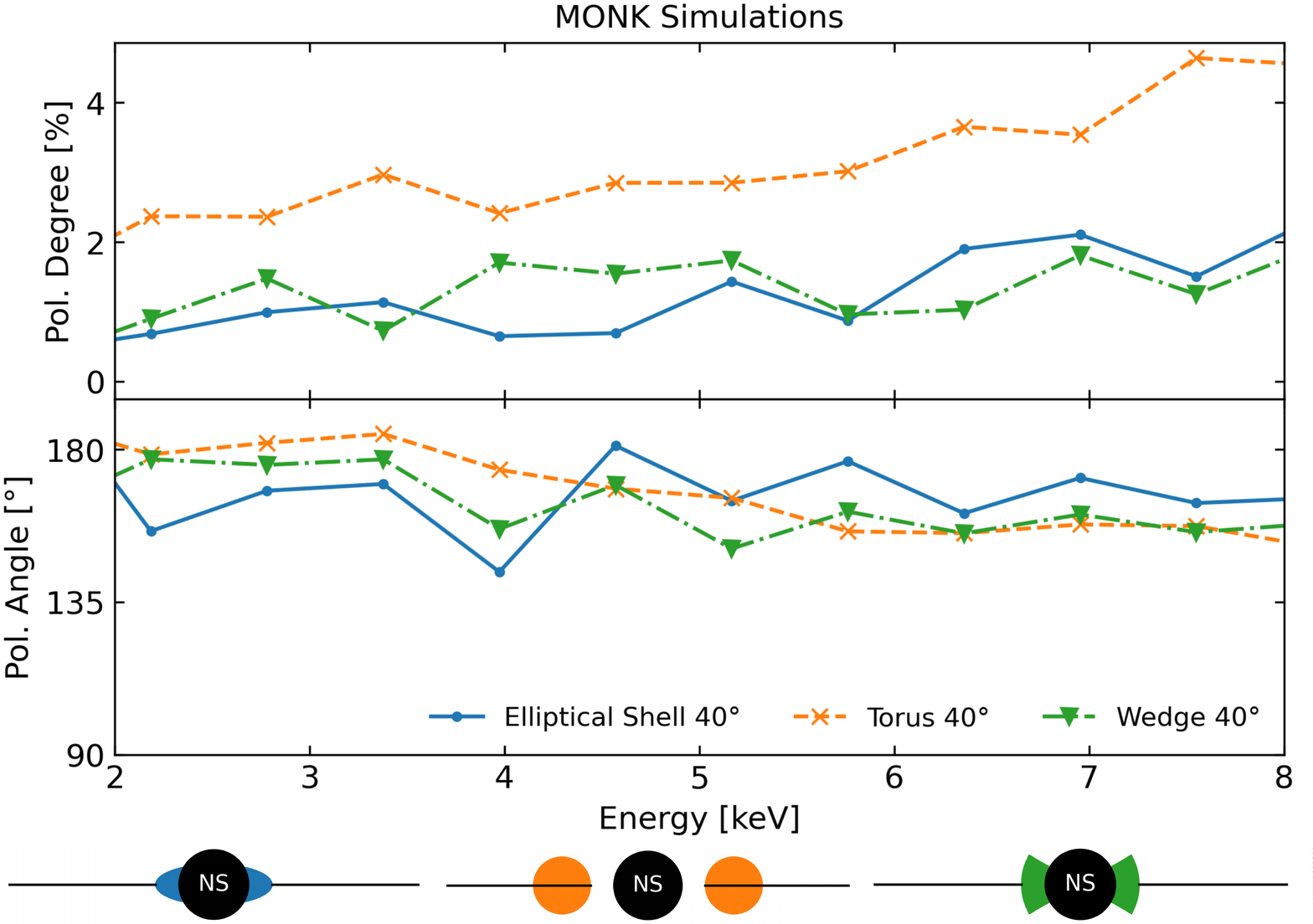}
\caption{
\textsc{monk} simulations of the PD and PA of the Comptonized component, as a function of the energy in the 2--8 keV band, for three geometries: elliptical shell, torus, wedge. The assumed inclination is 40\degr\ in all cases. The PA in \textsc{monk} is measured from the projection of the rotation axis onto the sky plane. 
}
\label{fig:monk_sim}
\end{figure}

We show in Fig. \ref{fig:monk_sim} simulations of PD and PA of the Comptonized component, performed with the relativistic ray-tracing code \textsc{monk} \citep{zhang2019} applied to the case of neutron stars \citep[for the details, see][]{gnarini2022}. We assume three different geometries: an ellipsoidal shell around the NS equator, a torus covering the disc, and a wedge. 
The wedge and elliptical shell are chosen in order to reproduce the SL/BL geometries. The wedge is a spherical shell without polar caps, that radially extends from the NS surface up to the inner edge of the accretion disc, and covers the NS up to a latitude of 30\degr. The elliptical shell is characterized by a semi-major axis coinciding with the inner disk radius, and a maximum latitude of 30\degr.
The torus has a minor diameter of 10 gravitational radii and covers part of the disk starting from its inner edge; it thus physically represents a puffed-up inner region of the disc. For the SL/BL geometries, the optical depth is measured in the radial direction from the NS surface \citep{popham&sunyaev2001}, while for the torus it is proportional to the minor radius.
In all cases, we set an electron temperature of 3.4 keV, a seed photon temperature of 1.575 keV, and an optical depth of 5.1, corresponding to the best fit values reported above (see Table \ref{tab:Fit}). 

We obtain a PD in the 2--8 keV band of roughly 1--2\% for the SL/BL and around 3--4\% for the torus. The PA is approximately 180\degr\ for all the three geometries, which in \textsc{monk} corresponds to polarization parallel to the rotation axis. The torus geometry seems to produce a polarization signal consistent with our results. However, the BL or SL geometry is more realistic if we assume that the accretion flow extends down to the NS surface -- as expected for a source in the soft state.

All in all, it is likely that the observed polarization comes from a combination of different components. For example, a BL or SL  with an ellipsoidal or wedge-like geometry could be consistent with the observed polarization, if we include the contribution from polarized radiation reflected off the disc. On the other hand, a small or null PD of the Comptonized component, which would be expected for a shell-like geometry \citep{gnarini2022}, seems to require a highly polarized reflection component. However, we cannot rule out the presence of a further, highly ionized reflection component, whose spectral shape would be almost indistinguishable from the primary continuum. 
For example, a highly ionized disk with $\log (\xi$/erg cm \pers)=4 would produce an essentially featureless reflection component \citep{relxillns}. According to our results, the observed reflection component is not so highly ionized, but the reflecting medium also does not extend down to the ISCO. This leaves open the possibility that part of the reflection is due to a highly ionized innermost region of the disk, extended down to the ISCO. Unfortunately, the present data do not allow us to place strong constraints on this putative component, because it is highly degenerate with the primary continuum and with the lower-ionization reflection component (which is in any case needed to produce the iron line).  If we include in our fits a further \relxillns\ component with $\log (\xi$/erg cm \pers)=4 and inner radius equal to the ISCO, its contribution to the 2--8 keV photon flux is between zero and 30\%. Therefore, we cannot exclude that ionized reflection from the innermost region of the disk could significantly contribute to the observed PD. 

As already discussed by \citet{farinelli2023} in the case of Cyg X-2, reflection is likely to have a significant impact on the polarization signal. On the other hand, if the reflection component was the only one to be polarized, it would require a very large PD. 
The PD of the reflected radiation is not obvious to predict because it depends on the assumed geometry, however, it is not likely to exceed $\sim$20\% \citep{matt1993,poutanen1996,schnittman&krolik2009}. 
It is worth mentioning the results of \citet{lapidus&sunyaev1985} who considered the case of radiation scattered off the accretion disk illuminated by the SL or the full NS surface corresponding to the emission during type I X-ray bursts. For an inclination angle $i \approx 40\degr$, the PD reaches $\approx$ 5\% for SL over NS radius to  height ratio $H/R_{\rm NS}=$0.1--0.2 (see Fig. 7 in their paper). As the authors did not consider the direct disk emission, it can be instructive to include it and perform a simple Stokes vectorial analysis. 
We define the two polarization pseudovectors (here in the form of normalized Stokes parameters) for the BL+reflection and direct disk  components:
\begin{equation} 
\begin{array}{l}
 q_i = P_i f_i \cos 2 \Psi_i ,\\
 u_i = P_i f_i \sin 2 \Psi_i ,
\end{array}
\end{equation}
where $P_i$ and $\Psi_i$ are the PD and PA of the two components ($i=1,2$), while $f_i$ is their relative contribution to the total photon flux in the 2--8 keV energy band. 
Using the values for $f_i$ reported in Table \ref{tab:Fit} and assuming  $P_1=5\%, \Psi_1=0\degr$ for BL plus reflection and $P_2=1\%, \Psi_2=90\degr$ for the disk 
\citep[i.e. parallel to the disk plane; see][]{farinelli2023}, 
the total PD and PA are 
${\rm PD}_{\rm tot}=(q_{\rm tot}^2+ u_{\rm tot}^2)^{1/2} = 1.94\%$ and ${\rm PA}_{\rm tot}=0\degr$,
with the value of ${\rm PD}_{\rm tot}$ being not far from the observed one. The same calculation in the 2--4 keV and 4--8 keV bands yields ${\rm PD}_{\rm tot}= 1.8\%$ and ${\rm PD}_{\rm tot}= 3.7\%$, respectively.
To produce the net total PD, the disk PD can be quite low, well consistent with the observed upper limit of 3\%--4\%. The energy-dependent PD value can be naturally explained by considering that the disk photon flux in the 2--4 keV energy range is about five times higher than that in the 4--8 keV interval, thus the effect of polarization cancellation by two components polarized at about right angles is less at higher energies. 
This analysis shows that the accretion geometry of GX~9+9, a bright atoll in the soft state whose spectrum resembles that of Z sources, is consistent with a BL/SL illuminating the ionized surface
of the disc, producing polarized reflection as well as the (unpolarized) fluorescent iron line.

If the high-energy components indeed dominate the X-ray polarization, a harder spectrum should be associated with larger polarization. Further observations of both atoll and Z sources in different spectral states will improve our understanding of the X-ray polarization of NS-LMXB, and thus of their accretion geometry.

\begin{acknowledgements}
We thank the anonymous referee for carefully reading the paper and for useful comments that improved the manuscript.
The Imaging X-ray Polarimetry Explorer (IXPE) is a joint US and Italian mission.  The US contribution is supported by the National Aeronautics and Space Administration (NASA) and led and managed by its Marshall Space Flight Center (MSFC), with industry partner Ball Aerospace (contract NNM15AA18C).  
The Italian contribution is supported by the Italian Space Agency (Agenzia Spaziale Italiana, ASI) through contract ASI-OHBI-2017-12-I.0, agreements ASI-INAF-2017-12-H0 and ASI-INFN-2017.13-H0, and its Space Science Data Center (SSDC) with agreements ASI-INAF-2022-14-HH.0 and ASI-INFN 2021-43-HH.0, and by the Istituto Nazionale di Astrofisica (INAF) and the Istituto Nazionale di Fisica Nucleare (INFN) in Italy.
This research used data products provided by the IXPE Team (MSFC, SSDC, INAF, and INFN) and distributed with additional software tools by the High-Energy Astrophysics Science Archive Research Center (HEASARC), at NASA Goddard Space Flight Center (GSFC).
JP and SST acknowledge support from the Academy of Finland grants 333112. 
POP acknowledges financial support from the French High Energy Programme of CNRS (PNHE) and from the French Space Agency (CNES).
\end{acknowledgements}


\bibliographystyle{aa}
\bibliography{bibliography-GX9p9}






\end{document}